\documentclass[12pt]{article}
\usepackage{graphicx}
\usepackage{amssymb}
\usepackage{latexsym}
\usepackage[dvips]{epsfig}

\newcommand{\pdrv}[2]{ \frac{\partial #1}{\partial #2}}

\newcommand{\pa}{\partial}

\newcommand{\be}{\begin{equation}}
\newcommand{\ee}{\end{equation}}
\newcommand{\bea}{\begin{eqnarray*}}
\newcommand{\eea}{\end{eqnarray*}}
\newcommand{\bean}{\begin{eqnarray}}
\newcommand{\eean}{\end{eqnarray}}
\newcommand{\overleftrightarrow}[1]{\vbox{\ialign{##\crcr
    $\leftrightarrow$\crcr\noalign{\kern-1pt\nointerlineskip}
    $\hfil\displaystyle{#1}\hfil$\crcr}}}

\newcommand{\td}[1]{\tilde{#1}}

\newcommand{\n}[1]{\label{#1}}

\begin{document}

\title{{\hfill {\small Alberta--Thy 17-02}}\vspace{2cm}\\
Capture and Critical Scattering of a Long Cosmic String by a
Rotating Black Hole} 
\author{Martin Snajdr\footnote{e-mail:
msnajdr@phys.ualberta.ca} ${}^1$ and Valeri Frolov\footnote{e-mail:
frolov@phys.ualberta.ca} ${}^1$}
\maketitle

\begin{center}
\noindent{
$^{1}${\em
Theoretical Physics Institute, Department of Physics, \ University of
Alberta, \\ Edmonton, Canada T6G 2J1}
}
\end{center}
\bigskip

\maketitle

\begin{abstract}      
The capture of a straight, infinitely
long cosmic string by a rotating black hole with rotation parameter $a$ is
considered. We assume that a string is moving with velocity $v$ and
that initially the  string is parallel to the axis of rotation of the
black hole and has the impact parameter $b$. The string can be either 
scattered or captured by the black hole.
We demonstrate that there exists a critical value of the impact 
parameter $b_c(v,a)$ which separates these two regimes.
Using numerical simulations we obtain the critical impact parameter
curve for different values of the rotation parameter $a$. We
show that for the prograde motion  of the string this curve
lies below the curve for the retrograde motion. Moreover, for 
ultrarelativistic strings moving in the prograde direction 
and nearly extremal black holes the critical impact parameter
curve is found to be a multiply valued function of $v$.
We obtain real time
profiles of the scattered strings in the regime close to the
critical. We also study critical scattering and capture of strings by
the rotating black hole in the relativistic and ultrarelativistic
regime and especially such relativistic effects as coil formation and
wrapping effect. 
\end{abstract}

\vspace{3cm}

\newpage
\section{Introduction}
\setcounter{equation}0

Cosmic strings are topologically stable one-dimensional objects which
are predicted by unified theories and can be created  during a phase 
transition in the early Universe \cite{ShVi:94}.  Recent measurements
of CMB anisotropy gave certain restrictions on the models of cosmic
string formation but do not exclude their existence. The analysis
shows that a mixture of inflation and topological defects is
consistent with current CMB data
\cite{PoVa:99,Cant:00,SiPe:01,LaSh:02,BoPeRiSa:02}.

When a cosmic string in its motion passes close to a black hole it
can be captured. In this paper we study this phenomenon. We consider
the string to be a test object moving in the background gravitational
field of a rotating black hole. Test string approximation can be
justified since the dimensionless parameter $\mu^*=G\mu/c^2$ ($\mu$
is the string mass per a unit length) is very small. For GUT energy
strings $\mu_{GUT}^*\sim 10^{-6}$ and for EW energy scale
$\mu^*_{EW}\sim 10^{-34}$. The emission of gravitational waves by the
string, which might be of astrophysical interest, affects the string
motion only in the order $\sim {\mu^*}^2$. This effect can be
considered by using the perturbation methods. But first one must
solve the equations of motion for the test string\footnote{In a
realistic setup there is always matter surrounding a black hole. 
The presence of matter results in a friction force acting on the 
moving string.
Calculations show (see e.g. \cite{GaWi:87}) that the energy lost by a
string segment of length $L$ during its motion through the region of
size $\Lambda$  containing matter of density $\rho$ can be estimated
as $\Delta E\sim \alpha G\mu\rho \Lambda^2Lv$, where $\alpha$
is a dimensionless parameter depending on the string's
$\gamma$-factor. The relative change of the string segment kinetic
energy $E=L\mu v\gamma c^2$ is ${\Delta E/E}\sim {\alpha G \rho
\Lambda^2/( \gamma c^2)}\, $. 
Denote $\rho_{BH}=M/R_g^3$, where $R_g$
is the gravitational radius of the black hole of mass $M$. Then
${\Delta E/ E}\sim  ({\rho/ \rho_{BH}})(\Lambda/ R_g)^2 \, $. Since this
quantity is very small we neglect the effect of interaction of a
string with the matter surrounding a black hole.}.  This problem by
itself is quite complicated because the equations are non-linear. If
a string remains far from the black hole, one can linearize the
string equations of motion. The solution of the linearized equations
in the black hole spacetime was obtained and studied in
\cite{DVFr:98,Page:98,Page:99,SnFrDV:02}. For the scattering at
smaller impact parameters, and especially for the string capture
non-linear effects are very important and the problem is to be
treated numerically. The string scattering and capture by a
non-rotating black hole was studied in \cite{DVFr:97,DVFr:99}. One
might expect that in the most of the astrophysically interesting
cases (such as a stellar mass black hole in a binary system or
massive and supermassive black holes in the center of galaxies and
quasars) the black hole is rotating or even to be close to extremally
rotating. String scattering by a rotating black hole  (including the
non-linear regime) was discussed in our previous paper
\cite{SnFrDV:02}.
We demonstrated that as a result of scattering
of an initially infinitely long straight string moving with 
velocity $v$ and an impact parameter $b$ the string, after passing 
the black hole, keeps moving with the same velocity, while the central part of the
string moves in a new plane with a different impact parameter.
The region of this `new phase' expands as the kink-like transition layers
connecting the `new' and `old' phases propagate with the velocity of light away
from the center of the string. 
We studied the string scattering data, i.e., the change of the impact
parameter and the width of the transition layers as a function of $v$ and
$b$. We showed that for large $b$ the effects of the rotation are
negligibly small and one can use a weak field approximation in this
regime. For strings coming closer to the black hole the effect of
rotation becomes more profound. We used numerical simulations to
obtain the string scattering data in different regimes of scattering
and the real time profiles of the strings.

\begin{figure}[ht]
\begin{center}
\epsfig{file=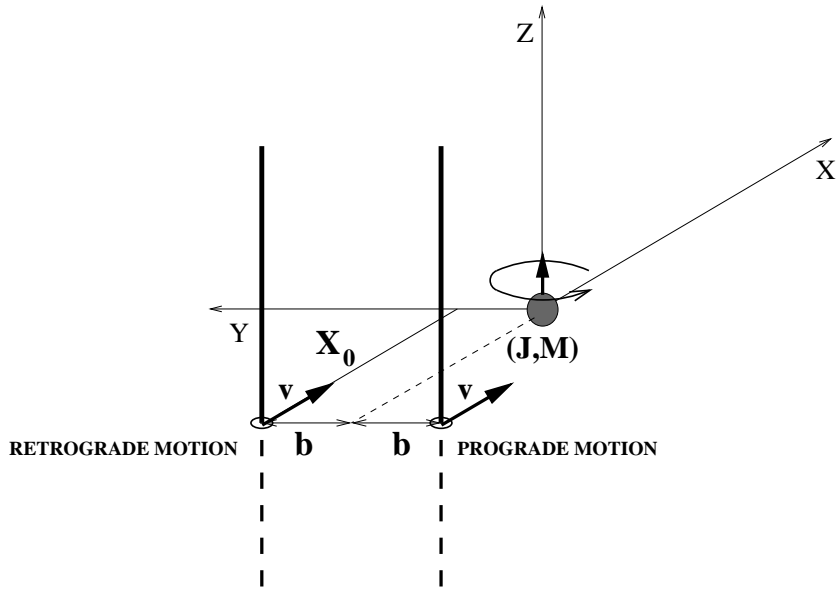, width=11cm}
\caption{Initial setup for prograde and retrograde string motion. The part of the
string depicted by the dashed line lies below the $X$-$Y$ plane.}
\label{setup}
\end{center}
\end{figure}

In this paper we continue studying an interaction of a long cosmic
string with a rotating black hole. Now we focus our attention on the
capture and scattering in the special regime close to the capture. 
We consider a black hole
of mass $M$ and angular momentum $J=aM$. We assume that initially a
straight infinitely long cosmic string is parallel to the axis of
rotation of the black hole, and it is moving with velocity $v$ and
impact parameter $b$. 
For a rotating black hole we distinguish two principal type of
motion.
If the  angular momentum of the black hole coincides 
with the direction of the angular momentum of the string
the motion is called a {\em prograde} motion.
In the opposite case it is called a {\em retrograde} motion.
The initial setup of the cosmic string is schematically shown on
figure \ref{setup}.
For large impact parameters, $b\gg M$, the
string after coming to the black hole moves further and escapes to
infinity.  The capture occurs when  the central part of the string
comes close to the black hole and eventually  crosses the event horizon.
By central part we will always understand the part of the string close
 to the $Z=0$ symmetry plane.

For a given velocity $v$ there exist a critical value of the impact
parameter, $b_{c}(v,a)$, which separates capture and scattering
regimes. 
As we shall show, scattering of the string with the impact
parameter slightly greater than $b_{c}(v,a)$ has a number of
interesting features. We call this regime {\em critical scattering}.
The second aim of this paper is to study this regime.

The paper is organized as follows. We remind the string equation of
motion in section~\ref{EOM}. In this section we also discuss the initial data
for our problem. String capture and critical impact parameter curves
are discussed in section~\ref{captureandcrit}. In section~\ref{nearcrit}
 for we present the real time
profiles for critical string scattering and discuss the form of these
profiles at the late time regime. The effect of coil formation
for the relativistic strings is considered in section~\ref{coilformation}.
 In this section we present the results of the numerical simulations for the
coil formation regions and compare them with an analytical
solution for the weak field scattering. The last section~\ref{discussion}
 contains general discussion and further remarks.

\section{String equation of motion}
\label{EOM}
\setcounter{equation}0

We consider long string motion in the Kerr spacetime with metric
\[
ds^2 = -(1-\frac{2Mr}{\Sigma})d\td{v}^2 + 2d\td{v}dr - 
         \frac{4aMr\sin^2{\theta}}{\Sigma}d\td{v}d\td{\phi} -
         a\sin^2{\theta}drd\td{\phi}
\]	 
\be \n{2.1}
+ \Sigma d\theta^2 + \frac{A\sin^2{\theta}}{\Sigma}d\td{\phi}^2 \, .
\ee
Here
\[
\Sigma = r^2+a^2\cos^2{\theta}\, ,\hspace{0.5cm}
\Delta = r^2 - 2Mr + a^2\, ,
\]
\be\n{2.2}
A = (r^2+a^2)^2 - a^2\Delta^2\sin^2{\theta} \ .
\ee
We use the so called Kerr  (in-going) coordinates
$(\td{v},r,\theta,\td{\phi})$ which are related to the standard
 Boyer-Lindquist coordinates $(t,r,\theta,\phi)$ as follows
\[
\td{v} = t+r+M\ln|{\Delta\over 4M^2}| + \frac{M^2}{\sqrt{M^2-a^2}}\ln
      \left|\frac{r-M-\sqrt{M^2-a^2}}{r-M+\sqrt{M^2-a^2}}\right| + M\,
      ,
\]
\be\n{2.3}     
\td{\phi} = \phi+\frac{a}{2\sqrt{M^2-a^2}}\ln
         \left|\frac{r-M-\sqrt{M^2-a^2}}{r-M+\sqrt{M^2-a^2}}\right|\ .
\ee

The string worldsheet $X(\zeta^A)$ ($\zeta^A$, $A=0,1$) obeys the following
equations
\be\n{2.4} 
\Box {\cal X}^{\mu}+h^{AB}\Gamma^{\mu}_{\alpha\beta}{\cal
X}^{\alpha}_{,A}{\cal X}^{\beta}_{,B}=0\, ,
\ee
\be\n{2.5} 
\gamma_{AB}-{1\over 2}h_{AB}h^{CD}\gamma_{CD}=0 \, ,
\ee
where
\be\n{2.6} 
\Box ={1\over \sqrt{-h}}\partial_A(\sqrt{-h}h^{AB}\partial_B)\, .
\ee
The first of these equations is the dynamical equation for string
motion, while the second one plays the role of  constraints. 
These equations provide an
extremum to the Nambu-Goto action (taken in the Polyakov form)
\be\n{2.7} 
I[{\cal X}^{\mu},h_{AB}]=-{\mu \over 2}\,\int d^2\zeta
\sqrt{-h}h^{AB}\gamma_{AB}\, .
\ee
We use units in which $G=c=1$, and the sign conventions of \cite{MTW}. In 
(\ref{2.7}) $h_{AB}$ is the auxiliary metric
with determinant $h$.

We fix internal coordinate freedom by using the gauge in which
$h_{AB}$ is conformal to the flat two-dimensional metric
 $\eta_{AB}=\mbox{diag}(-1,1)$.
In this gauge
the equations of motion for the string have the form 
\be\n{2.8}
 \Box_0 {\cal X}^\mu + \Gamma^\mu_{\alpha\beta}\, {\cal
 X}^{\alpha}_{,A}\, {\cal X}^{\beta}_{,B}\, \eta^{AB}\, =0\, ,
\ee 
and the constraints are
\be\n{2.9}
\gamma_{01} = g_{\mu\nu}\pdrv{{\cal X}^\mu}{\tau}\pdrv{{\cal X}^\nu}{\sigma} = 0\ 
,
\ee
\be\n{2.10}
\gamma_{00}+\gamma_{11} = g_{\mu\nu}\left(\pdrv{{\cal X}^\mu}{\tau}
\pdrv{{\cal X}^\nu}{\tau}+\pdrv{{\cal X}^\mu}{\sigma}\pdrv{{\cal X}^\nu}{\sigma}
\right)\ = 0\ .
\ee
Here, $\tau\equiv\zeta^0$, $\sigma\equiv\zeta^1$ and 
$\Box_0=-\pa^2_\tau+\pa^2_\sigma$.

It is easy to check that a solution for a straight string moving in
a {\em flat} spacetime can be written as follows
\be\n{2.11}
{\cal X}^{\mu}={\cal X}_0^{\mu}(\tau,\sigma)\equiv (\tau\cosh\beta,
\tau\sinh\beta+X_0, b,\sigma) \, ,
\ee
\be\n{2.12} 
h_{AB}=\eta_{AB}\equiv \mbox{diag}(-1,1)\, .
\ee
This
solution describes a straight string oriented along the $Z$-axis
which moves in the $X$-direction with constant velocity $v=\tanh
\beta$. Initially, at ${\tau}_{0} = 0$, the string is found at 
${\cal X}^{\mu}(0,\sigma) = (0,X_0, b,\sigma)$. 
We call $b$ an {\em impact parameter}. 
In our simulations we choose $b$ to be positive and 
$X_0$ to be large and negative.

To study string scattering and capture by the black hole we use
numerical simulation. The general scheme is described in details in 
\cite{SnFrDV:02}. However, the simulation of critical scattering is 
computationally more involved and the adaptive mesh refinement 
algorithm has been improved. In our simulation we alter the grid density
according to a truncation error estimate. If the error estimate 
in certain region of the grid lies above
a prescribed threshold the calculation is restarted from a last stored 
state with the grid 
in the region (and its neighborhood) appropriately refined.
Similarly, if the truncation error in some region of the grid
 is much smaller than the threshold 
the grid in that region is coarsened (this time, of course,
  it is not necessary to restart the calculation).

Here we just make some general remarks concerning
the calculations. We assume that initially the string is very far
away from the black hole where the spacetime is practically flat. We
use the solution (\ref{2.11})) in this region. When the
string comes closer to the black hole the gravitational field of the
latter modifies its motion. When the distance $r$ from the black hole
is much larger than the gravitational radius, $r>>M$, the
gravitational field is still weak. One can linearize the string equations
and obtain {\em linear} equations describing the string motion. It is
possible to show (see  \cite{SnFrDV:02}) that in this regime the
effects connected with the rotation of the black hole can be
neglected with respect to the {\em Newtonian} corrections. By solving
the weak field equations one determines the string evolution in the
weak field region. This solution is used to obtain initial data for
numerical simulations. It also can be used to get the boundary
conditions at ``numerical ends'' of the string. During the
simulations the dynamical equations are numerically solved while the
constraint equations are used as an additional check of the accuracy 
of the calculations.

\section{String capture and critical impact parameter}
\label{captureandcrit}
\setcounter{equation}0

\begin{figure}[ht]
\begin{center}
\epsfig{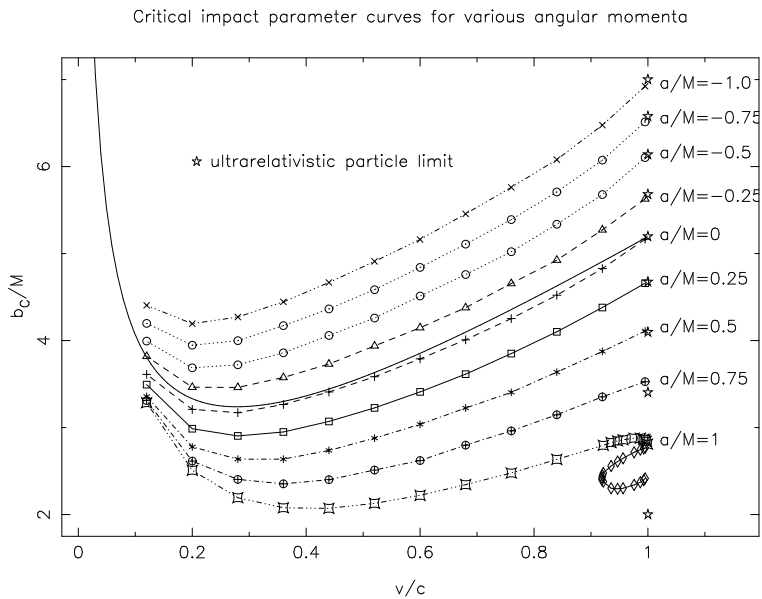}
\caption{Critical impact parameter $b_c$ as a function of the string's
initial velocity $v$}
\label{capt}
\end{center}
\end{figure}

Two outcomes are possible for the string evolution---scattering and
capture. Capture occurs when a part of the string enters the event
horizon of the black hole located at 
\be\n{2.13} 
r=M+\sqrt{M^2-a^2}\, . 
\ee 
Since the Kerr in-going coordinates are regular at the event
horizon we can trace the string evolution both in the black hole
exterior and interior. For our study of the string capture we stop
our calculations as soon as part of the string crosses the
horizon.
The value of $b_c(v,a)$ is determined as follows. For each fixed velocity
$v$ and angular momentum $a$ we start 
the search with two values of $b$---one for which the string is captured
and one for which it escapes. Then we use bisection method to 
bracket the critical impact parameter\footnote{This algorithm must be used
with caution when the critical impact parameter curve contains multiple branches.}.
The results of
these calculations are presented at figure~\ref{capt}.
The error bars are smaller than the data markers so they are not shown.

This plot contains the critical impact parameter curves for 9
different values of the rotation parameter $a$ from $a=-M$  to
$a=M$. 
The signs are chosen so that a negative value of $a$ corresponds to 
a retrograde motion of the string and positive value  to a
prograde one. The lower the value of $a$ the higher is the position
of the critical impact parameter curves in the
$(v,b_c)$--plane. This feature is easily explained by the {\em
dragging into rotation} effect. Indeed, for the retrograde motion the
dragging effect slows down the string's motion. The string spends more time in
the strong attractive field of the black hole and hence it can be captured
easier. In order to escape the capture the string must be moving with
larger impact parameter. Thus the critical impact parameter is larger
that for a non-rotating black hole. For the prograde motion the effect
is opposite.

Numerical calculations were done for $v/c>0.12$. For smaller velocity
the impact parameter becomes large and computational time grows. On
the other hand, for large impact parameters the main part of the time
evolution of the string occurs in the weak field region. One can use
this to estimate the critical impact parameter for low velocity
motion. It was done by Don Page \cite{Page:98}. He
proposed
the following approximation for the critical  impact parameter
\be\n{2.14}
b_c/M = \left(\sqrt{\frac{\pi}{2v}} - \sqrt{\frac{\pi}{2}}\right) 
-\frac{64}{15}\, (1-v) +3\sqrt{3}\, . 
\ee
This function is shown in figure~\ref{capt} by a solid
line. One can see that it gives quite good approximation for a
non-rotating black hole.
For rotating black holes this line gives good
approximation for small $v$ since for large impact parameters
the rotation plays less important role. 

It is instructive to compare the string scattering with the
scattering of particles. In particular, in the ultrarelativistic limit
$v/c \to \infty$ the critical impact parameter for the
string coincides with the critical impact parameter for photons. The
only exception is a case of prograde scattering by a nearly extreme
black hole. Complicated structure of the critical impact parameter
curve for this case is connected with the features of the near
critical scattering which we discuss later.
The capture parameter for particles in the the ultrarelativistic limit
is (see e.g. \cite{FrNo:98})
\be\n{2.15}
b^{part}_{c}/M = 8\cos^3\left[\frac{1}{3}(\pi-\arccos a)\right]
+a\, .
\ee
The value of $b^{part}_{c}$ for $a/M=0, \pm 0.25, \pm 0.5, \pm 0.75,
\pm 1$
is shown by stars on figure~\ref{capt} on the vertical line at $v/c=1$. These
points are very close to those belonging to an ultrarelativistic string.
The only exception is a region of $a/M$ close to 1.

\begin{figure}[ht]
\begin{center}
\epsfig{file=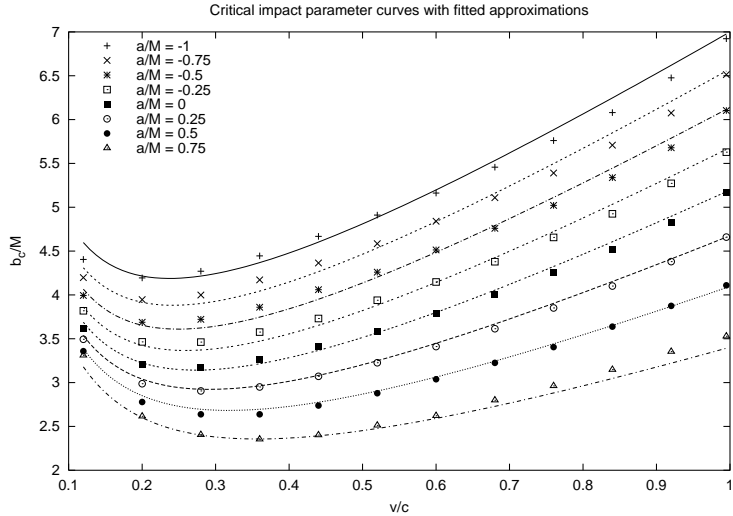, width=10cm}
\caption{Critical impact parameter $b_c$ and its fitted approximation
$b^{fit}_c$ as  functions of the string's
initial velocity $v$ for different values of the rotation parameter
$a/M$.}
\label{fit}
\end{center}
\end{figure}

The Page's approximation for critical impact parameter curves can be generalized to the
case of a rotating black hole. We can write the corresponding ansatz
in the form
\be\n{2.16}
b^{fit}_c/M = \left(\sqrt{\frac{\pi}{2v}} - \sqrt{\frac{\pi}{2}}\right) 
-(b_0+b_1\, a +b_2\, a^2)\, (1-v) +b^{part}_{c}/M\, . 
\ee
The best fit for numerically calculated critical impact parameter curves gives the
following values of the fitting parameters $b_i$
\be\n{2.17}
b_0= 4.40\, ,\hspace{1cm}
b_1= - 1.55\, ,\hspace{1cm}
b_2= -0.53\, .
\ee
Figure~\ref{fit} shows the result of the fitting for different values of
$a/M=0, \pm 0.25, \pm 0.5, \pm 0.75, -1$. (We exclude $a/M=1$ because
of the peculiar behavior of the critical impact parameter curve for this case.) It is
easy to see that continuous curves representing $b^{fit}_c$ given by
(\ref{2.16})--(\ref{2.17}) are in a good agreement with the numerical
data shown by shaped points.

\begin{figure}[ht]
\begin{center}
\epsfig{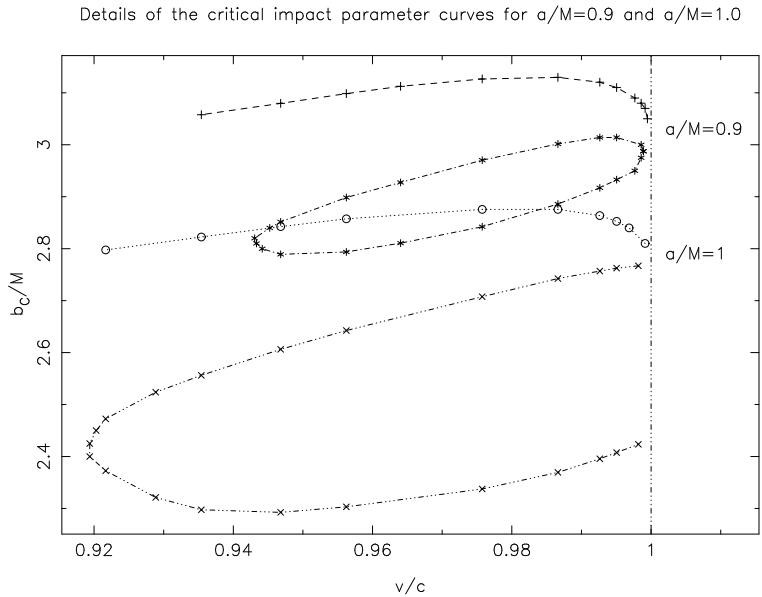}
\caption{Capture curve for a prograde scattering of 
a relativistic cosmic string by
a rapidly rotating black hole.}
\label{cr_capt}
\end{center}
\end{figure}

Let us discuss an additional new feature in the critical impact curves
which occurs for highly relativistic prograde scattering by rapidly
rotating black holes. Plot for $a/M=1.0$ presented in figure~\ref{capt} shows that in
this regime $b_c(v,a)$ becomes a multiply valued function of $v$. We
studied numerically this regime. Figure~\ref{cr_capt} shows the
corresponding region in more details. It contains plots of $b_c(v,a)$
near $v/c=1$ for two values of the rotation parameter---$a/M=0.9$ and
$a/M=1.0$.
Besides the main branch curve (marked by `$+$' for $a/M=0.9$ and by `$\circ$'
for $a/M=1.0$) there also exists an additional branch 
(marked by `$\ast$' for $a/M=0.9$ and by `$\times$' for $a/M=1.0$).
We performed calculations up to $v/c=0.9995$ which
corresponds to the gamma-factor $\gamma\approx 30$. 
For greater values of $\gamma$ our program does not allow to obtain the solution with
the required accuracy.

For given velocity $v$  lying just below the main branch the string
is captured. But if the value of impact parameter lies inside the closed loop,
the string is again able to escape to infinity. 
We call this peculiar
behavior of near-critical strings the {\em escape phenomenon}. At first
glance this behavior looks quite strange. We discuss the mechanism
which makes these types of motion possible in the next section.

\section{Near critical  scattering}
\label{nearcrit}
\setcounter{equation}0

\subsection{Real time string profiles for near critical  scattering}

\begin{figure}[h!]
\center
\begin{tabular}{cc}
\epsfig{file=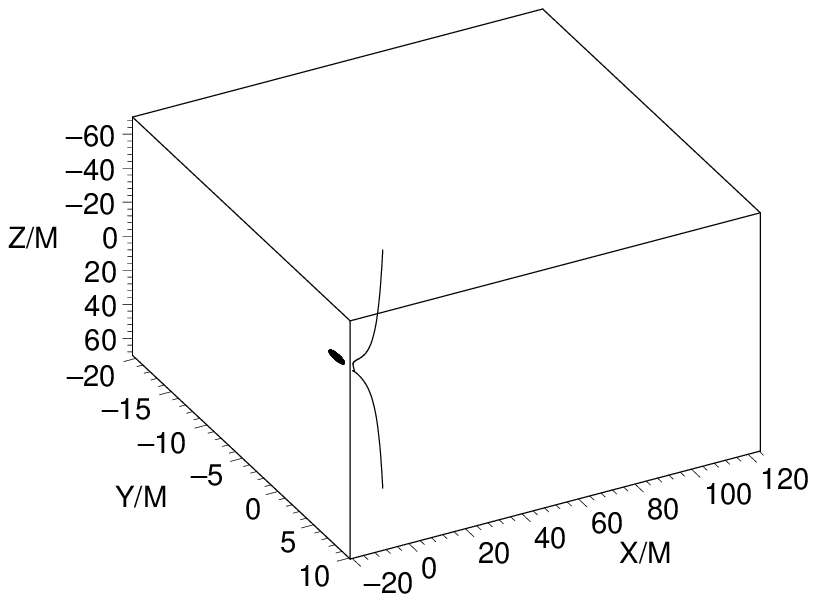, height=3.2cm} & \hspace{3cm}
\epsfig{file=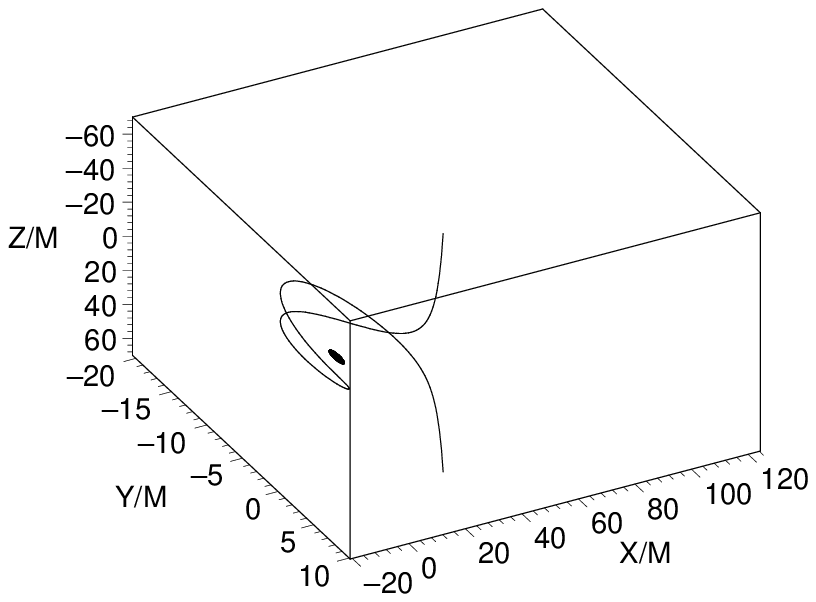, height=3.2cm}  \\
\vspace{0.5cm}
{\bf (a)} T/M=0& \hspace{3cm} {\bf (b)} T/M=21.88  \\
\epsfig{file=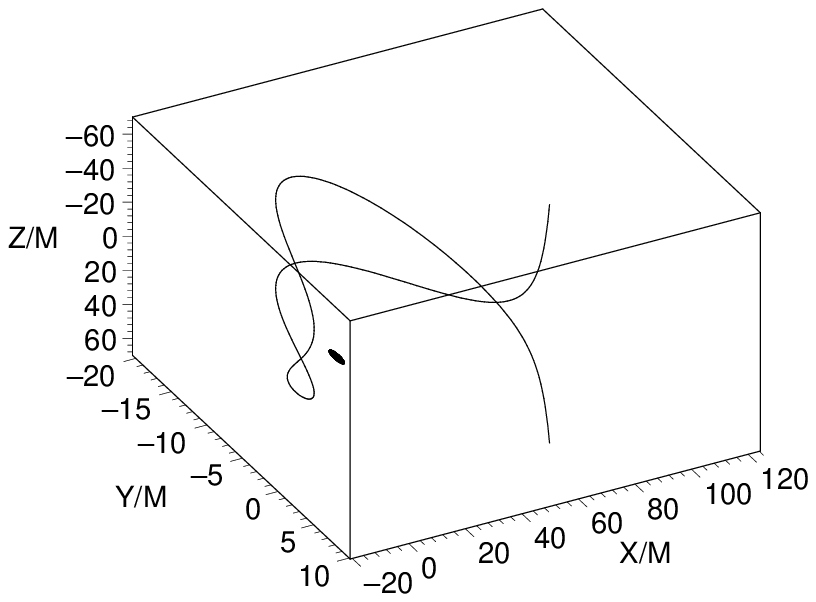, height=3.2cm} & \hspace{3cm}
\epsfig{file=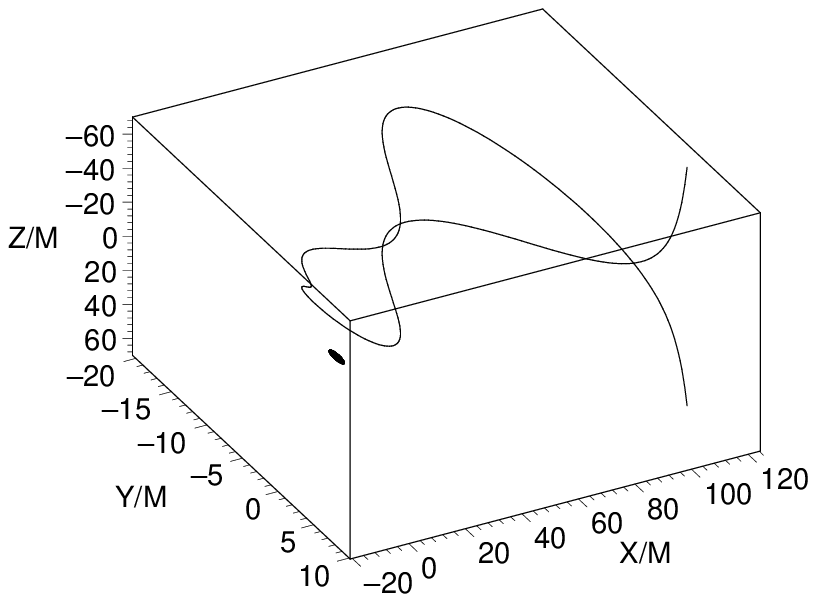, height=3.2cm}  \\
 {\bf (c)} T/M=41.31 & \hspace{3cm} {\bf (d)} T/M=60.51\\
\end{tabular}
\caption{Real time profiles of the cosmic string: $v/c=0.995$, $a/M=1$ and
$b/M=2.9$.}
\label{RT_1}
\bigskip
\begin{tabular}{cc}
\epsfig{file=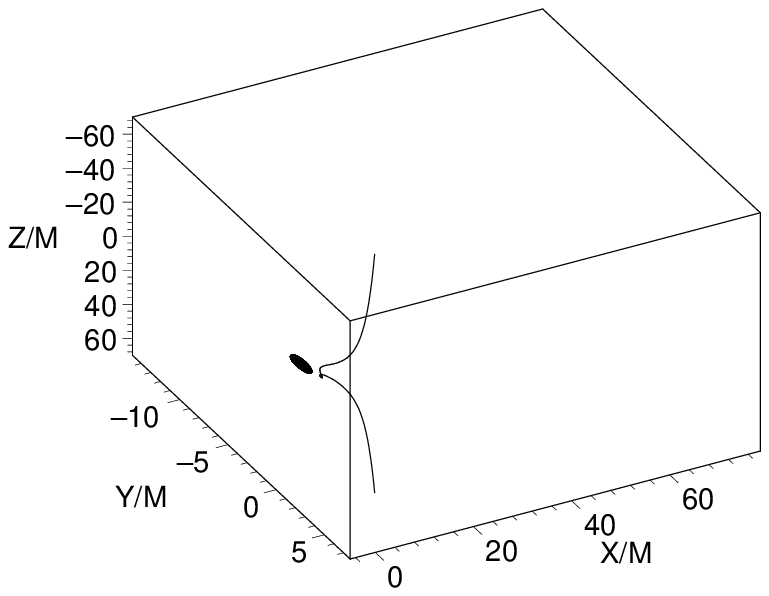, height=3.2cm} & \hspace{3cm}
\epsfig{file=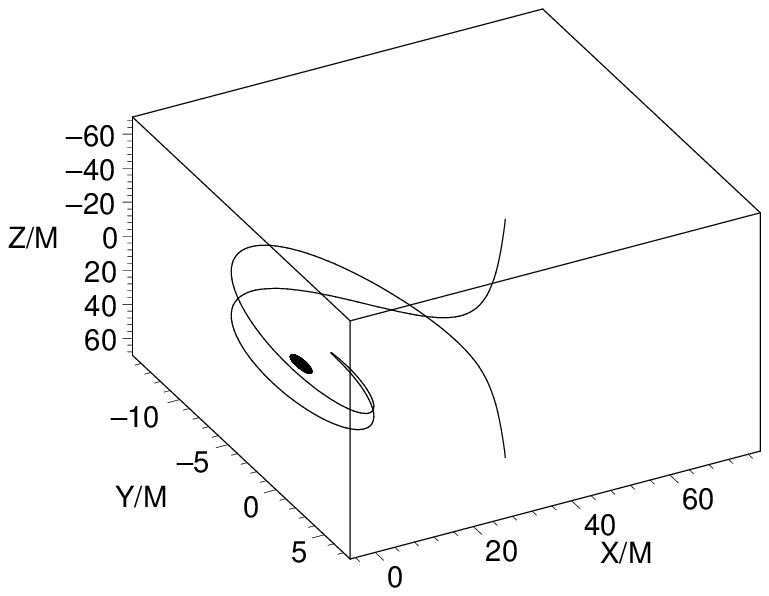, height=3.2cm} \\
\vspace{0.5cm}
{\bf (a)} T/M=0& \hspace{3cm} {\bf (b)} T/M=27.25\\
\epsfig{file=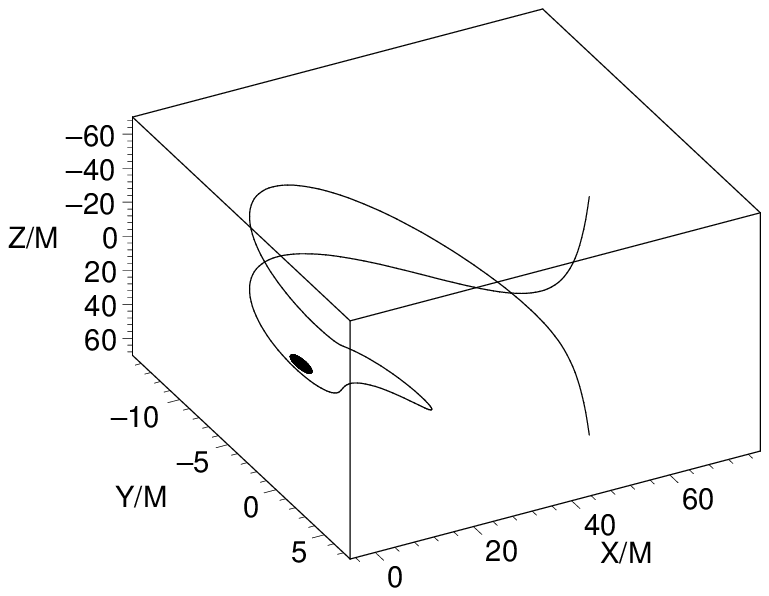, height=3.2cm} & \hspace{3cm}
\epsfig{file=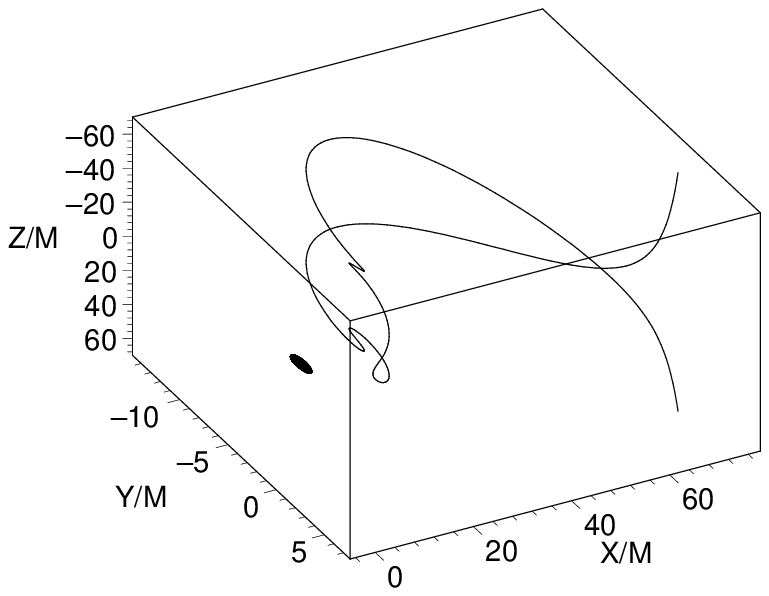, height=3.2cm} \hfill \\
{\bf (c)} T/M=44.76& \hspace{3cm} {\bf (d)} T/M=63.30\\
\end{tabular}
\caption{Real time profiles of the cosmic string: $v/c=0.995$, $a/M=1$ and
$b/M=2.55$.}
\label{RT_2}
\bigskip
\end{figure}

\begin{figure}[htb]
\center
\begin{tabular}{cc}
\epsfig{file=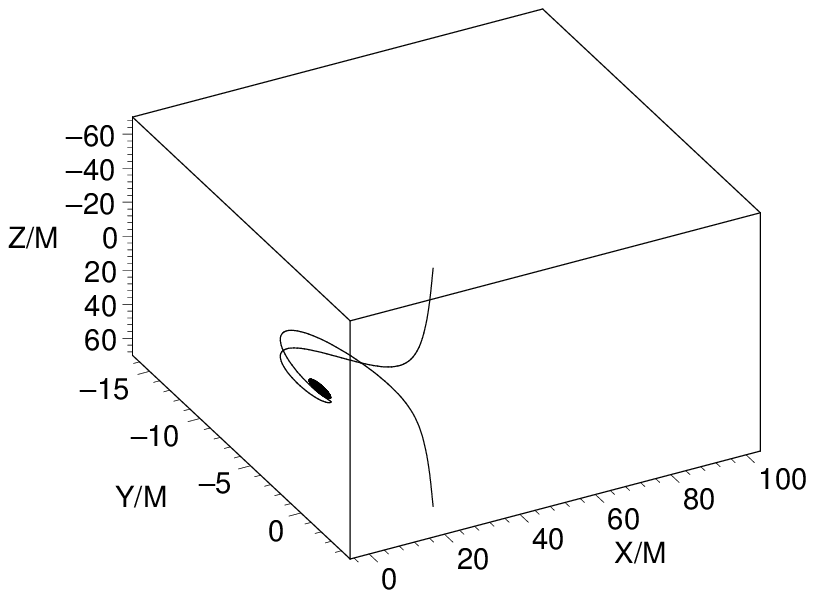, height=3.2cm} & \hspace{3cm}
\epsfig{file=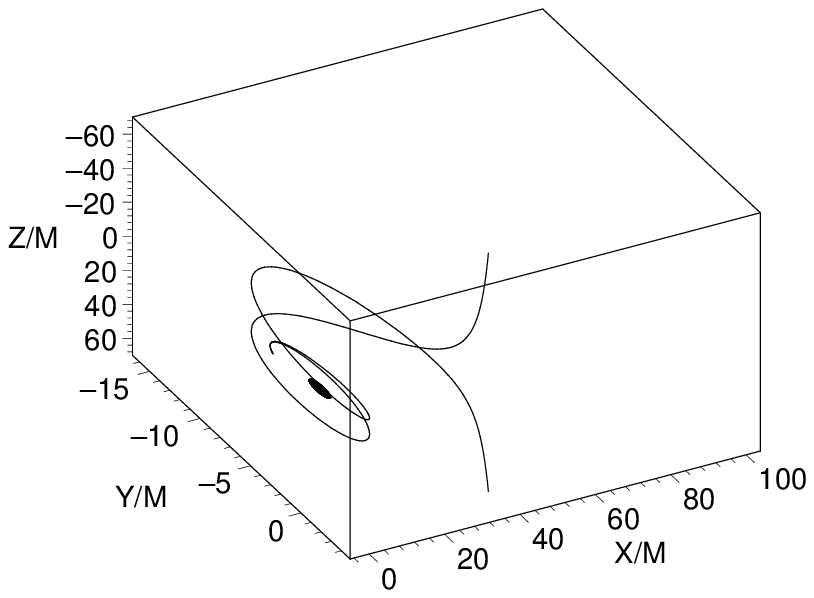, height=3.2cm} \\
\vspace{0.5cm}
{\bf (a)} T/M=15.13& \hspace{3cm} {\bf (b)} T/M=30.19\\
\epsfig{file=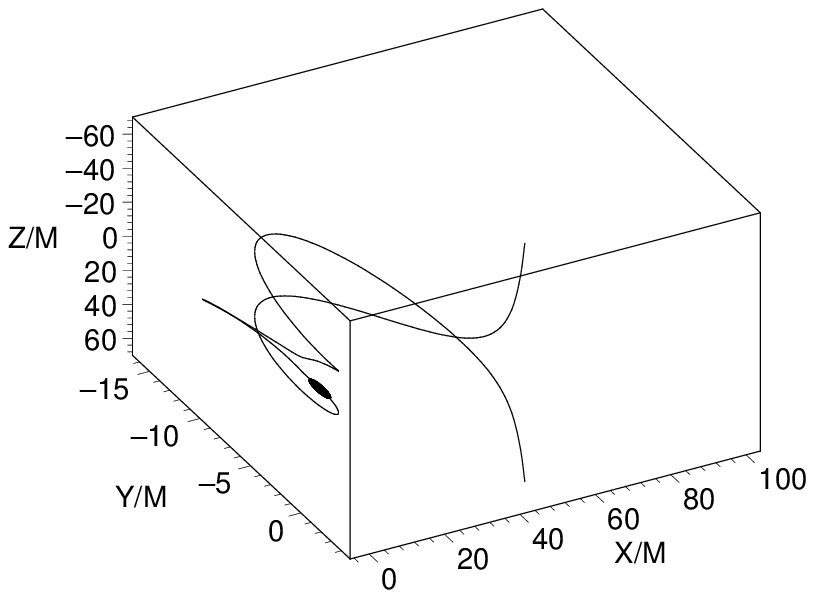, height=3.2cm} & \hspace{3cm}
\epsfig{file=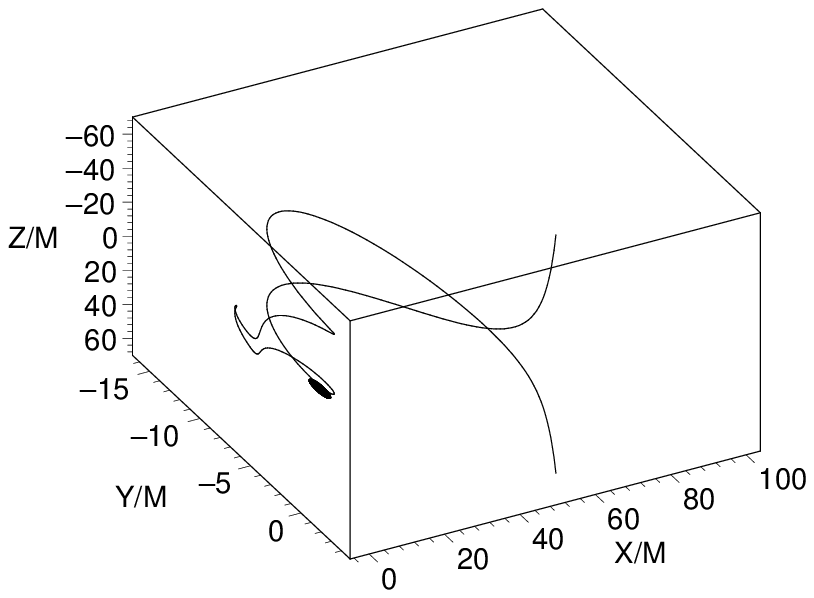, height=3.2cm} \\
\vspace{0.5cm}
{\bf (c)} T/M=40.10& \hspace{3cm} {\bf (d)} T/M=48.57\\
\epsfig{file=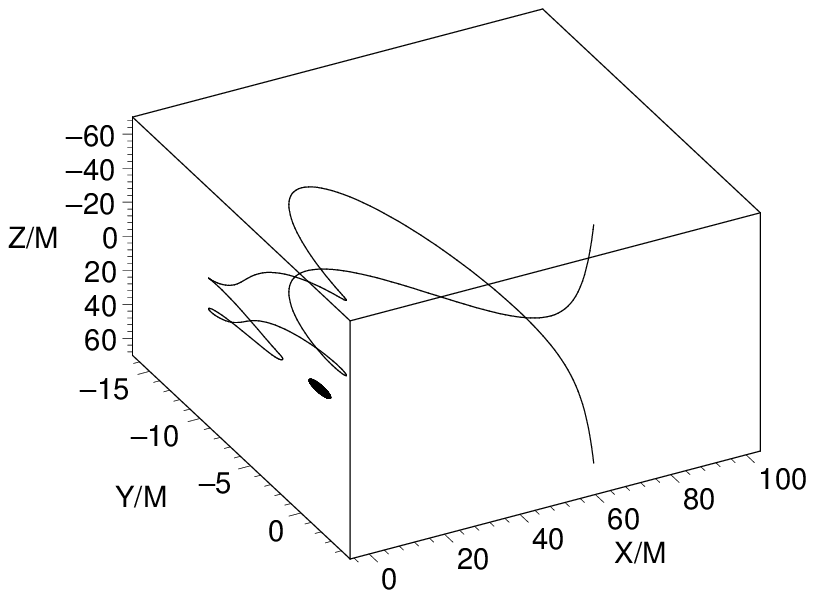, height=3.2cm} & \hspace{3cm}
\epsfig{file=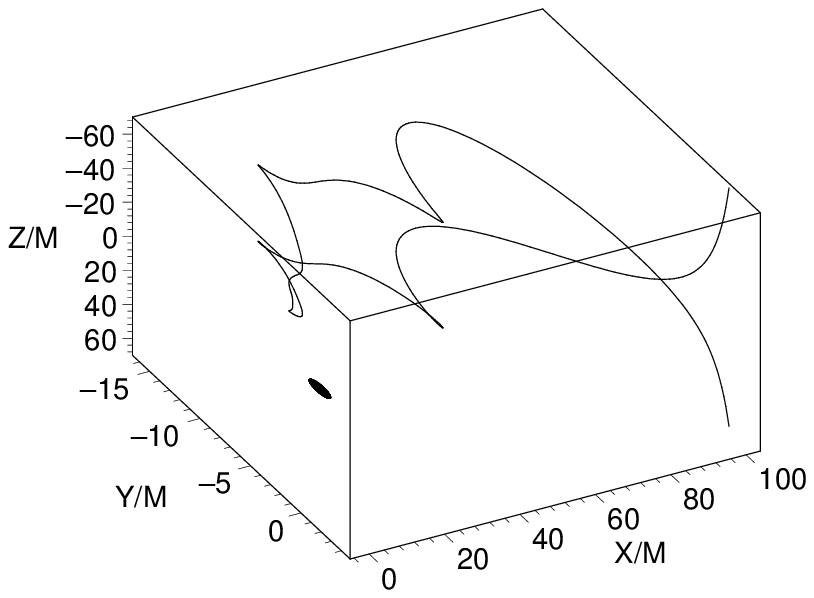, height=3.2cm} \\
{\bf (e)} T/M=58.87& \hspace{3cm} {\bf (f)} T/M=95.65\\
\end{tabular}
\caption{Real time profiles of the cosmic string: $v/c=0.995$, $a/M=1$ and
$b/M=2.4125$.}
\label{RT_3}
\end{figure}

As we saw in the previous section the critical impact parameter  curve is more
complicated in the  ultrarelativistic regime for a rapidly rotating
black hole. This can be better understood from the way the string is
scattered in the critical regime.
Figures~\ref{RT_1}--\ref{RT_3} illustrate this.
Each of the figures contains a set of snapshots made at some
moment of coordinate time $T$. We call such a snapshot a {\em real time profile}.
The time is given in units of $GM/c^3$.
$T=0$ corresponds to the moment when the parts of the string located far from 
the black hole (where the spacetime is practically flat) pass the $X=0$ plane.
The black hole event horizon is depicted as a black spot. It should be noted
that the scales along different axes are different. That is the reason 
why instead
of a round circle the spot representing the  black hole event horizon looks like
an ellipse.

There are two basic modes how a string
can escape. In the "standard" mode the string passes the black hole
with its the center partially wrapped around  it (figure~\ref{RT_1}, (a)--(b)).
As the rest of the string escapes the central part unwraps and the
string escapes to infinity (figure~\ref{RT_1}, (c)--(d)). As we lower
the impact parameter the central part wraps more around the black
hole and ultimately it does not have enough time to unwrap and it
gets captured. However, if we lower the impact parameter under
a certain value the string can escape again.
Now the mechanism is different. In this "alternative" mode the 
string wraps around the black hole as in the "standard" mode
(figure~\ref{RT_2}, (a)--(b)) but this time instead of unwrapping the black hole slips
through the open coil (figure~\ref{RT_2}, (c)--(d)). For even smaller 
values of the impact parameter this mechanism breaks
down eventually and the string is captured again. Although it is possible
that there exists another "island of escape" below the lowest
branch of the critical impact parameter curve we did not observe it in our simulations.

To have a better picture of the two different escape modes we created 3
MAPLE animations, each corresponding to one of the
figures~\ref{RT_1}--\ref{RT_3}.
These can be found at the URL http://www.phys.ualberta.ca/\~{}frolov/CSBH.
Note that the animations show only the central part of the string.
In the animations the rate of time is ``slowed down'' when the string
is close to the black hole in order to make the details of the string behavior more clear.
For this reason it seams that the string starts moving more rapidly when all its parts 
leave the proximity of the black hole.

Interesting features of the string structure are connected with {\em
wrapping effect}, that is, when the central part of the string rotates
around the black hole at the angle greater than $2\pi$. Such wrapping type of motion
is characteristic for the motion of the ultrarelativistic particles in
the regime close to capture. The tension of the string makes this
effect less profound. One can expect that for highly
ultrarelativistic motion it occurs even for a non-rotating black
hole. 
We do not see this effect in our simulations since we can not
perform sufficiently accurate calculations for ultrarelativistic
velocities beyond $\beta\approx 4.0$.
The dragging on effect connected with the rotation of the black
hole makes this effect more profound.  Figure~\ref{RT_3} shows real
time profiles for the string which "wraps" around the black hole.
It is interesting that during this process sharp spikes are formed in
the string profiles\footnote{These spikes are not infinitely sharp however, 
so the appropriate derivatives are well defined.}.

\subsection{Late time scattering data}

\begin{figure}[h!]
\center
\begin{tabular}{cc}
\epsfig{file=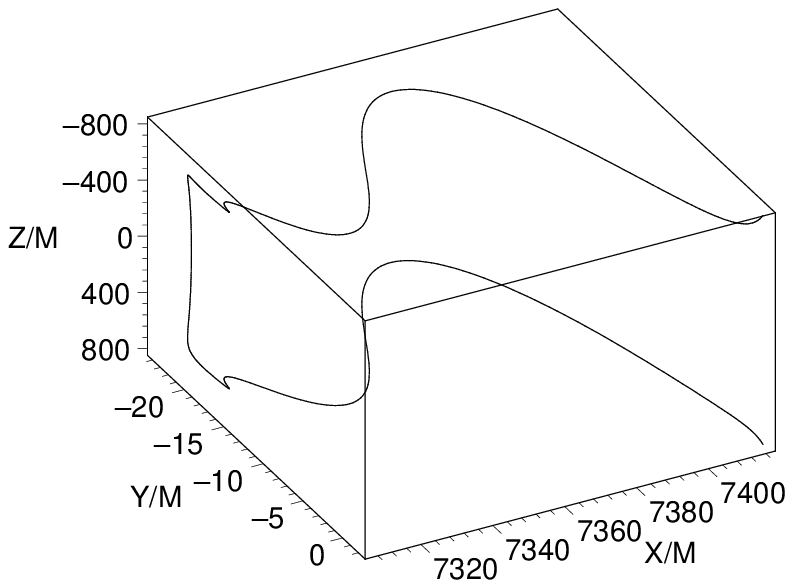, height=3.2cm} & \hspace{2.5cm}
\epsfig{file=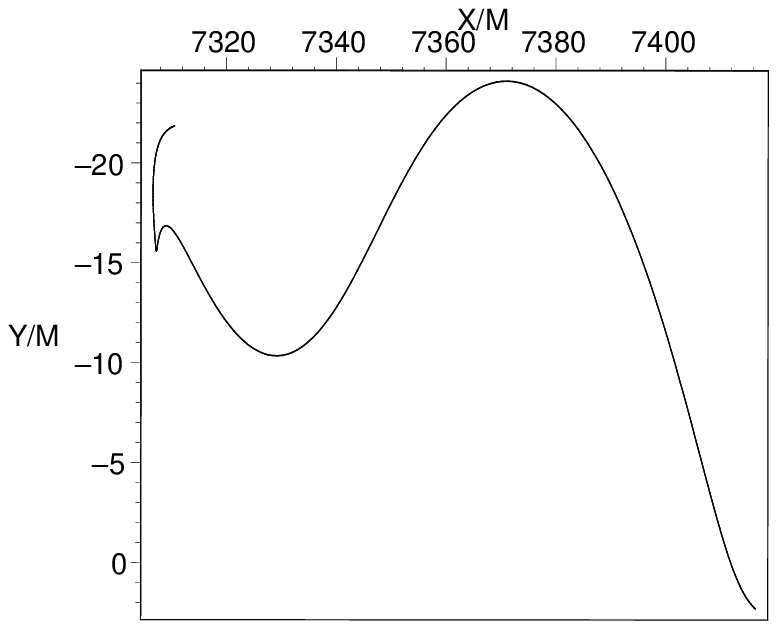, height=3.2cm} \\
\vspace{0.5cm}
{\bf (a)} & \hspace{2.5cm} {\bf (b)}\\
\epsfig{file=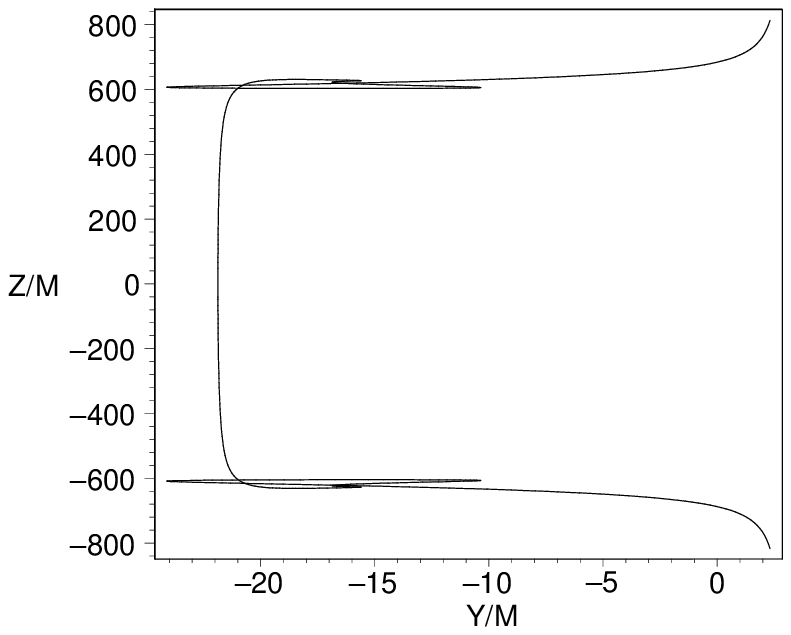, height=3.2cm} & \hspace{2.5cm}
\epsfig{file=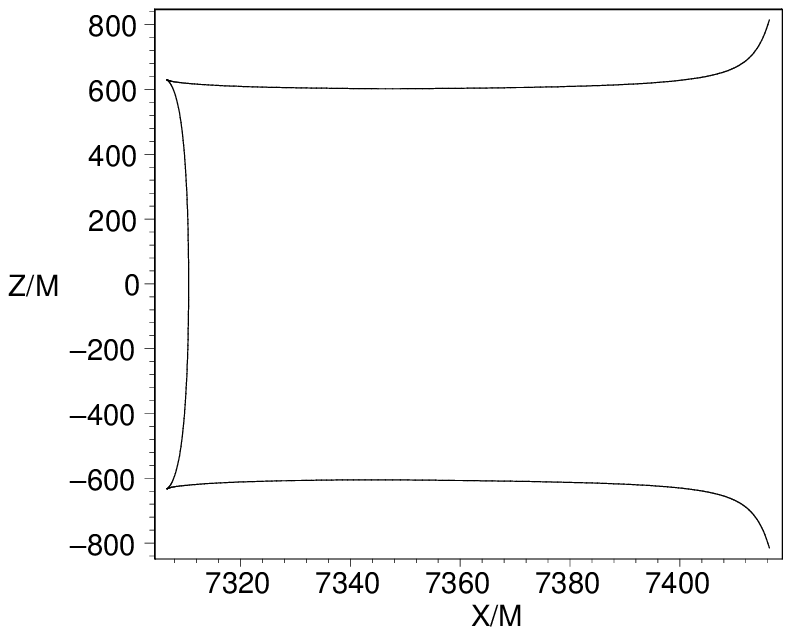, height=3.2cm} \\
{\bf (c)} & \hspace{2.5cm} {\bf (d)} \\
\end{tabular}
\caption{Late time profiles of the cosmic string: $v/c=0.995$, $a/M=1$ and
$b/M=2.9$, $T/M=7446.47$.}
\label{LT_1}
\bigskip
\begin{tabular}{cc}
\epsfig{file=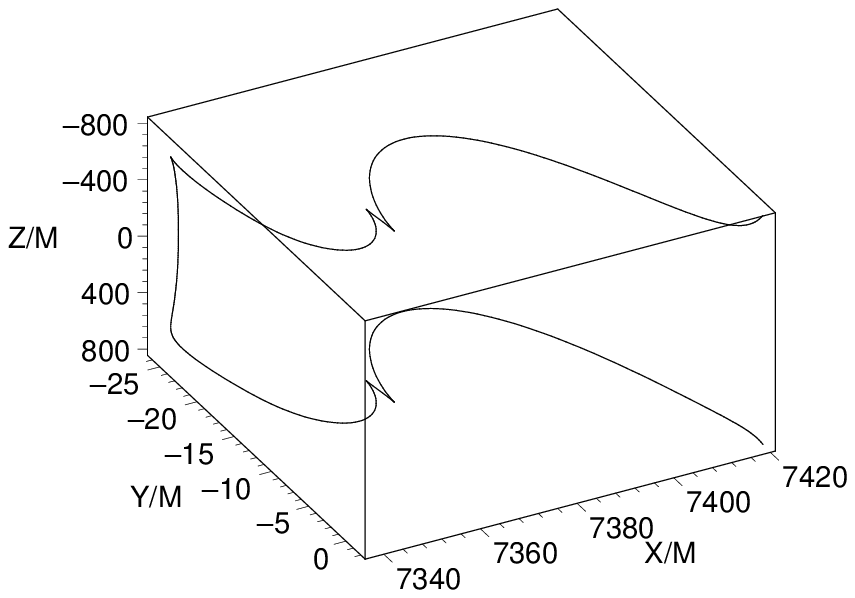, height=3.2cm} & \hspace{2.5cm}
\epsfig{file=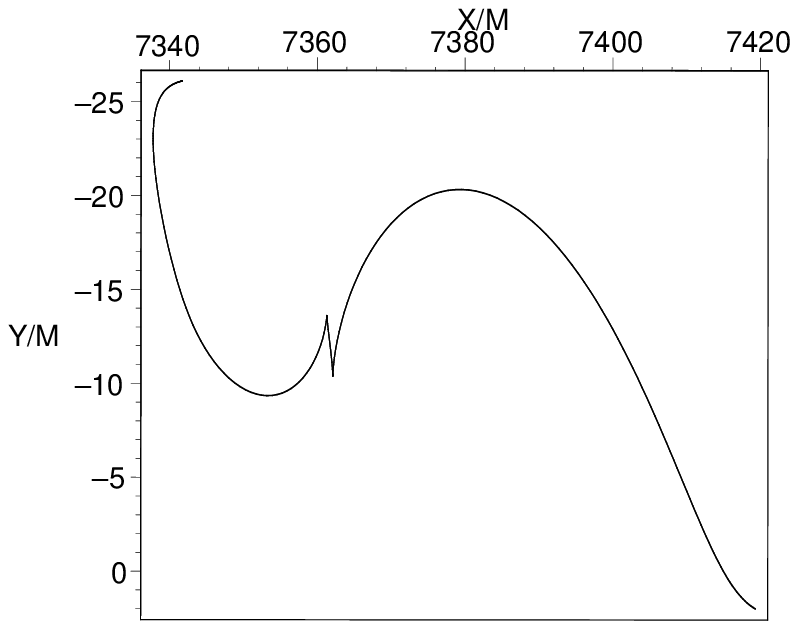, height=3.2cm}  \\ 
\vspace{0.5cm}
{\bf (a)} & \hspace{2.5cm} {\bf (b)}\\
\epsfig{file=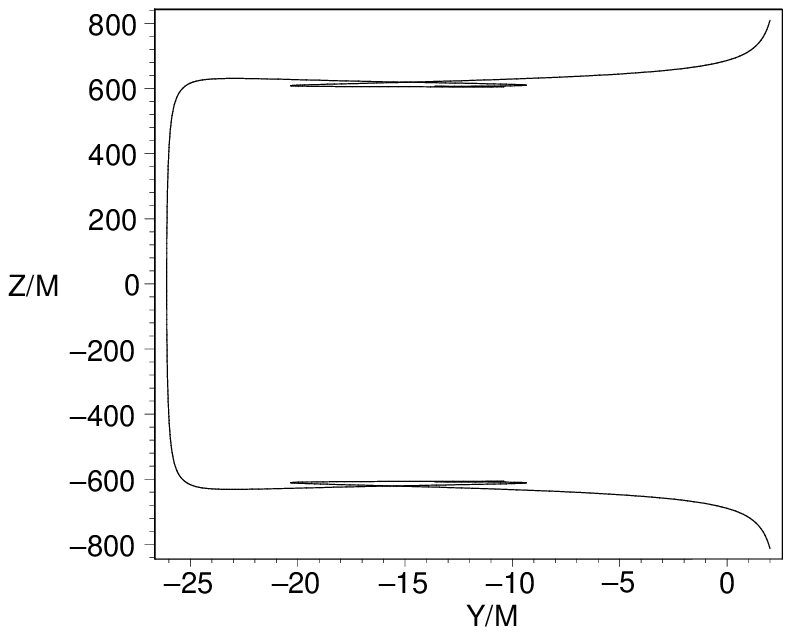, height=3.2cm} & \hspace{2.5cm}
\epsfig{file=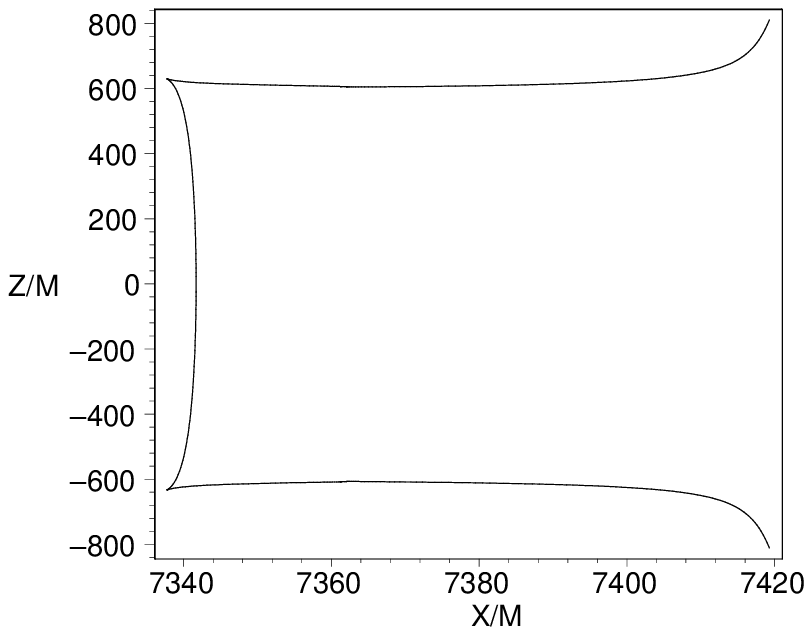, height=3.2cm} \\
{\bf (c)} & \hspace{2.5cm} {\bf (d)}\\
\end{tabular}
\caption{Late time profiles of the cosmic string: $v/c=0.995$, $a/M=1$ and
$b/M=2.55$, $T/M=7448.56$.}
\label{LT_2}
\end{figure}

\begin{figure}[h!]
\center
\begin{tabular}{cc}
\epsfig{file=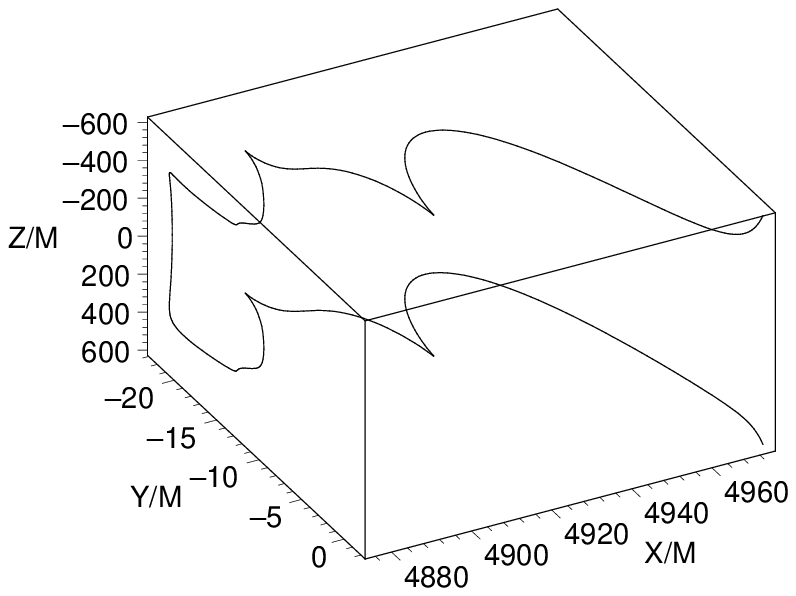, height=3.2cm} & \hspace{2.5cm}
\epsfig{file=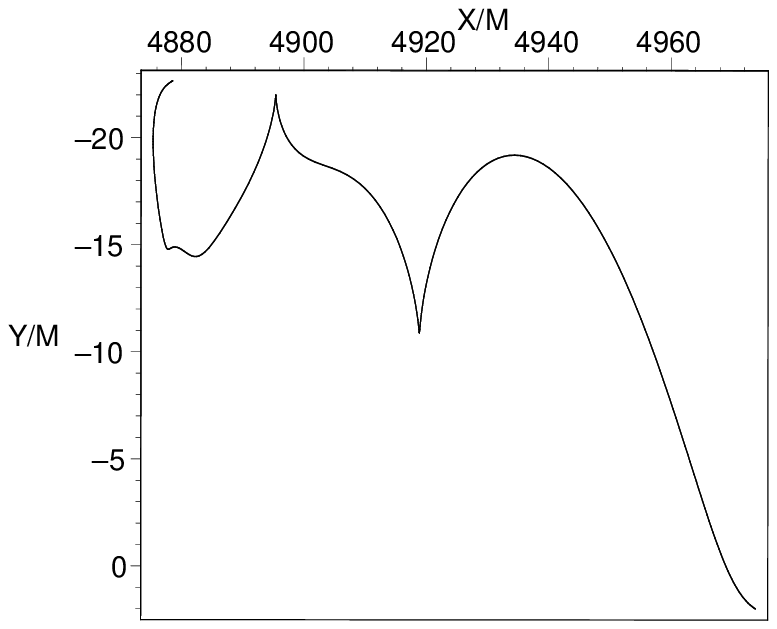, height=3.2cm} \\
\vspace{0.5cm}
{\bf (a)} & \hspace{2.5cm} {\bf (b)}\\
\epsfig{file=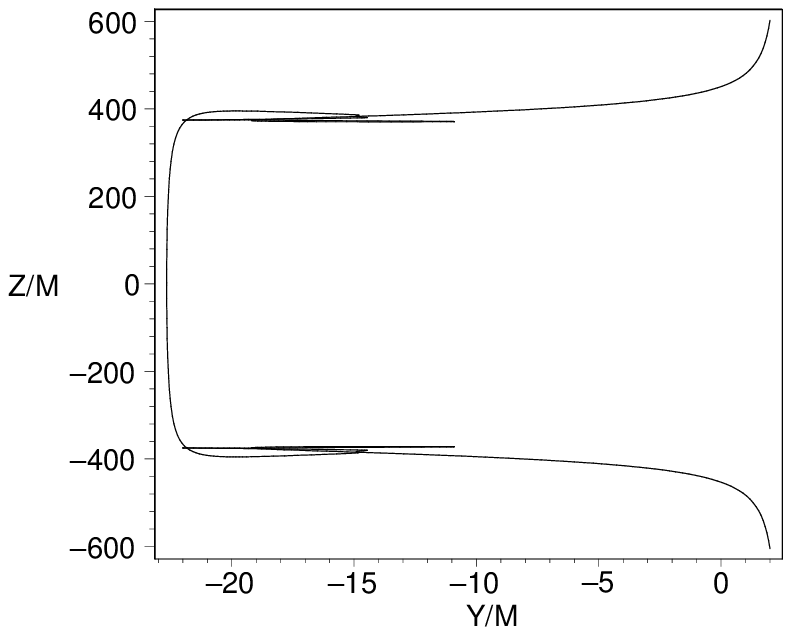, height=3.2cm} & \hspace{2.5cm}
\epsfig{file=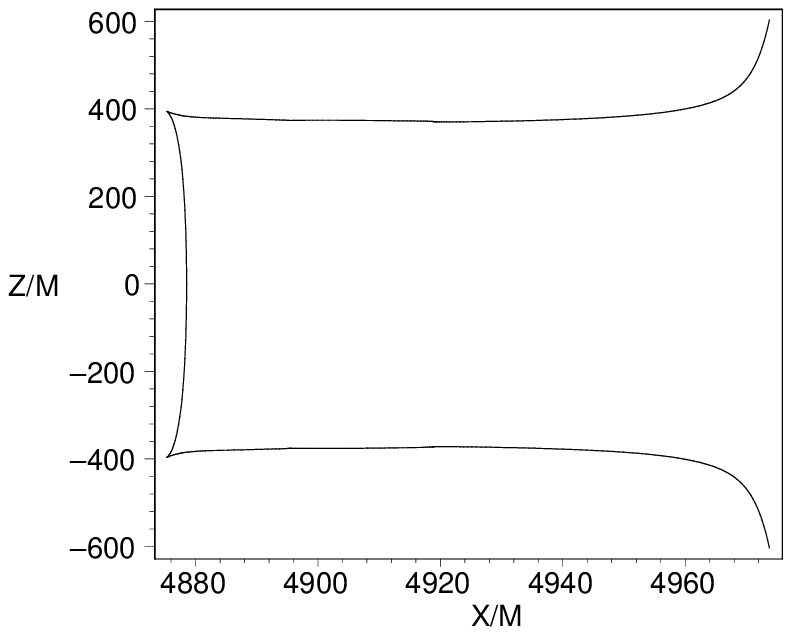, height=3.2cm} \\
{\bf (c)} & \hspace{2.5cm} {\bf (d)}\\
\end{tabular}
\caption{Late time profiles of the cosmic string: $v/c=0.995$, $a/M=1$ and
$b/M=2.4125$, $T/M=4991.53$.}
\label{LT_3}
\bigskip
\begin{tabular}{cc}
\epsfig{file=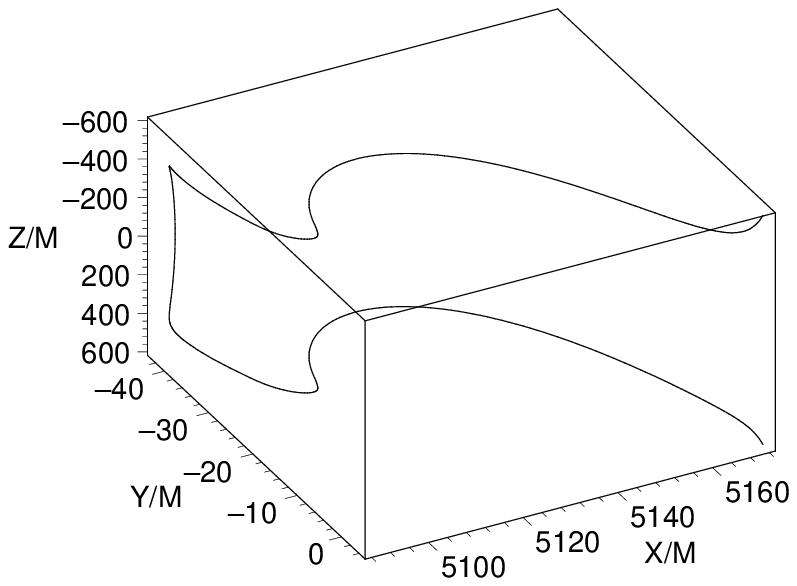, height=3.2cm} & \hspace{2.5cm}
\epsfig{file=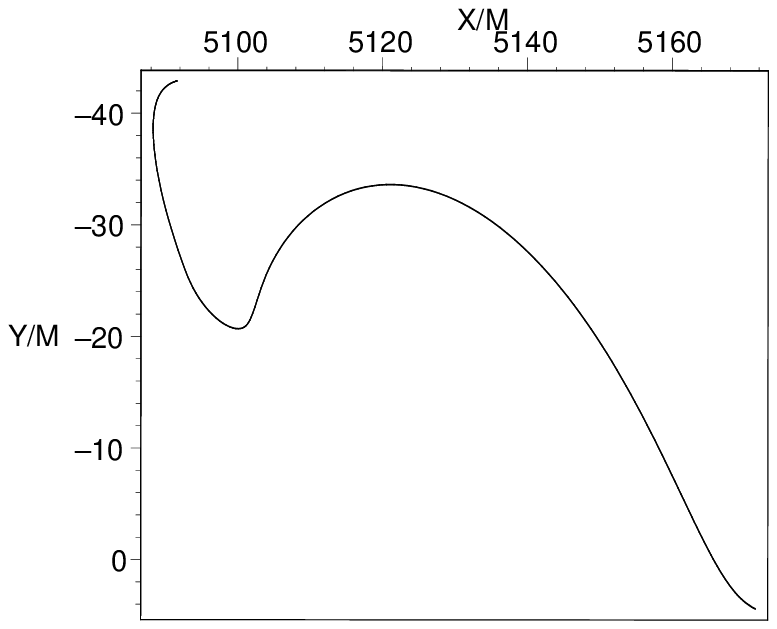, height=3.2cm} \\
\vspace{0.5cm}
{\bf (a)} & \hspace{2.5cm} {\bf (b)}\\
\epsfig{file=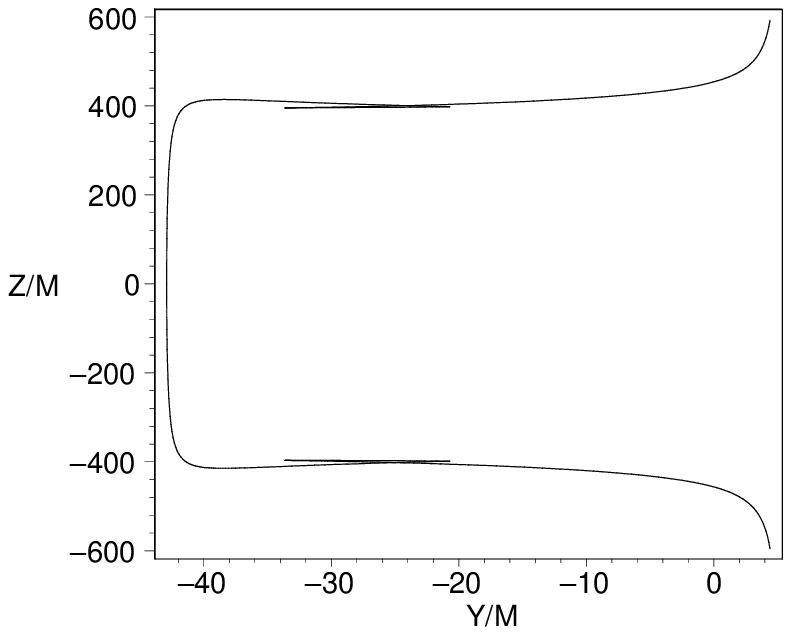, height=3.2cm} & \hspace{2.5cm}
\epsfig{file=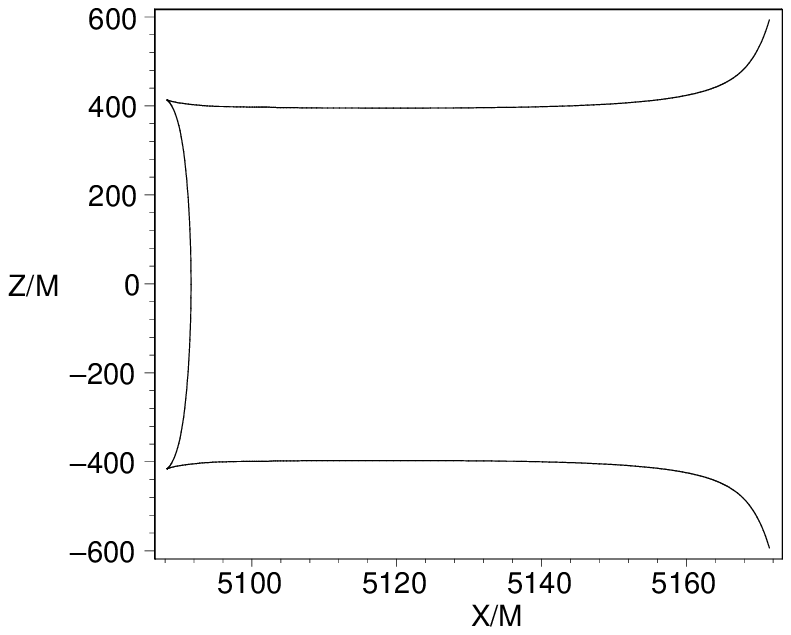, height=3.2cm} \\
{\bf (c)} & \hspace{2.5cm} {\bf (d)}\\
\end{tabular}
\caption{Late time profiles of the cosmic string: $v/c=0.995$, $a/M=0$ and
$b/M=5.17$, $T/M=5209.56$.}
\label{LT_4}
\end{figure}

For near critical scattering the central part of the string spends
some time in the vicinity of the black hole while the further located
parts of the string keep moving forward.  After the central part of the string
leaves the black hole vicinity the string as a whole is moving away
from the black hole with the central part excitations propagating along the
string. The combination of these effects results in quite a rich
structure of the late time string profiles.
Figures~\ref{LT_1}-\ref{LT_4} illustrate this.

Each of these figures consists of four plots. They demonstrate the form of
the string some time after it passes close to the black hole. The
black hole left behind the string is not shown. We show only that
part of the string near the center which contains interesting details.
The strings depicted on the figures \ref{LT_1}--\ref{LT_3} differ only by their 
initial impact parameter $b$. They are scattered by an extremal black hole in
the prograde direction.
For comparison figure \ref{LT_4} shows the late time profiles of a critically scattered string
by a Schwarzschild black hole.
Plots (a) show the real time 3-dimensional space profile of the
string, while plots (b), (c), and (d) show 2-dimensional projections
onto the planes $X$-$Y$, $Y$-$Z$, and $X$-$Z$, respectively. One can easily
observe that at this relatively late time when the interaction with
the black hole becomes weak the part of the string close to the
center looks practically as a segment of a straight line. This property
is also characteristic for a generic (non-critical)
string scattering. Namely, it was demonstrated that for this case as a
result of scattering the central part of the string `shifts' into the $Y$-direction and
a new `phase' is formed. This new phase is that region of the string which is moving in
the plane parallel to the initial one, but is displaced by some value
$\Delta Y$ in the $Y$-direction. 
The length of the string segment in the new phase grows with the velocity of light, the
transition layers being kink-like profiles. It is remarkable that a
similar picture is valid also in the critical regime, as one can see
by comparing plots (c) at figures \ref{LT_1}--\ref{LT_4}. An important
new feature is the possibility that the points with maximum shift in
$Y$ direction may be located not at the center of the string, but
slightly aside. Figure~\ref{LT_1} (c) shows that there are two such
points located symmetrically with respect to the center. 

The $X$-$Z$--profiles (plots (d) at figures
\ref{LT_1}-\ref{LT_4}) are also quite regular and resemble similar
profiles for a non-critical scattering.
On the contrary, the $X$-$Y$--profiles (plots (b) at figures
\ref{LT_1}-\ref{LT_4})  are rich of details. The spikes in the
string shape are most profound in this projection. All these details
are connected with a simple fact that the central part of the string spends
considerable time moving near the black hole, while the other parts of
the string are moving away practically with the velocity of light.
The profiles are sharper for the strings scattered by a rotating black
holes.  
One can expect that by fine tuning the impact parameter one might be
able to obtain string configurations which remain `glued' to the
vicinity of the black hole for arbitrary long time. But one can also
expect that such solutions are very unstable, so that a tiny change
of the impact parameter either results in the capture of the string or
in its earlier escape. This situation is similar to the one discussed
in \cite{FrLa:99}.  The paper demonstrated that axially symmetric
motion of a  circular string in the gravitational field of
Schwarzschild black hole is chaotic. One can make a conjecture that
the near-critical scattering of the cosmic string is also chaotic.

\section{Coil formation}
\label{coilformation}
\setcounter{equation}0

\begin{figure}[htb]
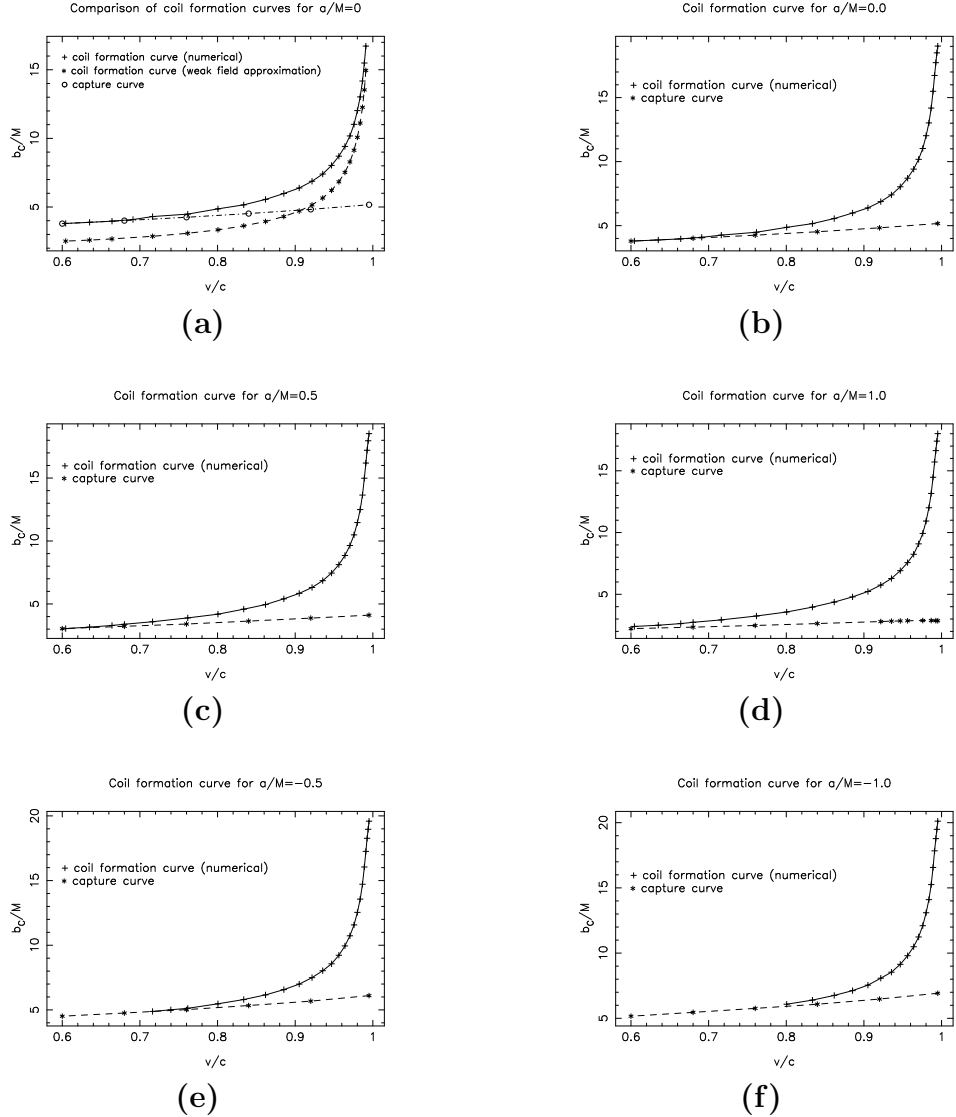

\center
\begin{tabular}{cc}
\epsfig{file=loop_comp.eps, width=5cm} & \hspace{2cm}
\epsfig{file=loop_0p0.eps, width=5cm} \\
\vspace{0.3cm}
{\bf (a)} & \hspace{2cm} {\bf (b)} \vspace{0.3cm}\\
\epsfig{file=loop_0p5.eps, width=5cm}& \hspace{2cm}
\epsfig{file=loop_1p0.eps, width=5cm} \\
\vspace{0.3cm}
{\bf (c)} & \hspace{2cm} {\bf (d)} \vspace{0.3cm} \\
\epsfig{file=loop_-0p5.eps, width=5cm}& \hspace{2cm}
\epsfig{file=loop_-1p0.eps, width=5cm} \\
{\bf (e)} & \hspace{2cm} {\bf (f)} \vspace{0.3cm} \\
\end{tabular}
\caption{Coil formation regions for prograde and retrograde scattering
of the string.}
\label{coil}
\end{figure}

There exists a special regime of the string scattering when after
passing nearby a black hole the string  forms a {\em coil}. A coil
arises when the function $Z(\tau,\sigma)$ looses its property to be a
monotonous function of $\sigma$ at fixed $\tau_0$. In this case there
exist two different values $\sigma_0$ and $-\sigma_0$ for which the
value of $Z$ is zero, $Z(\tau_0,\sigma_0)=Z(\tau_0,-\sigma_0)=0$.
Since all other coordinates are symmetric functions of $\sigma$, the
point $(\tau_0,\pm\sigma_0)$ is in fact a point of the string self
intersection. A coil exists for some finite time interval from
$\tau_i$ till $\tau_f$. At the end points of this interval
$\partial_{\sigma}Z|_{\sigma=0}=0$.

It should be emphasized that, because of the interconnection effect,
for most of the string models any  self-crossing of the string results in
the formation of a loop. After this, the loop moves independently from
the remaining part of the string. In our simulations we did not
include this effect. For this reason, such  details as a coil size and
structure may not represent the real properties of the string after
its self-intersection. On the other hand, before the intersection
occurs the numerical simulation describes the correct picture. That is
why we focus our attention on the form of the {\em coil formation
curve}. This is a curve (or surface) in the space of parameters which
separates the region without coils from the region with coils. The
structure of the coil formation curve does not depend on the details
of the loop formation process.

Figure~\ref{coil} shows the regions of coil formation for different
values of the rotation parameter $a$. This region lies below the 
coil formation curves shown in the pictures. From below this region
is restricted by the critical impact parameter curve. For a point lying exactly on the coil
formation curve one has $\tau_i = \tau_f=\tau_*$. It means that
$\partial_{\sigma}Z|_{\sigma=0}$ has a local minimum at
this point as a function of $\tau$ and hence
\be\n{}
\partial_{\sigma}Z|_{(\sigma=0,\tau=\tau_*)}=
\partial_{\tau}\partial_{\sigma}Z|_{(\sigma=0,\tau=\tau_*)}=0\,
.
\ee

The plots presented in figure~\ref{coil} demonstrate that  the
formation of coils occurs when a string moves with relativistic
velocity. For the prograde motion the coil formation starts at
$v/c\approx 0.6$ (see figure~\ref{coil}, (c) and (d)). For the
retrograde motion the corresponding velocity is higher and reaches
$v/c\approx 0.8$ for the extremally rotating black hole (see
figure~\ref{coil}~(e) and (f)). 

It is instructive to compare these results with the condition of the
coil formation in the weak field approximation.  In order to do this
we need to solve the weak field equations (see \cite{DVFr:98}). We first
solve these equations for the case when a straight string at some
initial moment of time $\tau=0$ is located at $X=X_0$,  where $X_0$
is a large distance of the string from the black hole. We
obtain a unique solution of perturbed equations by assuming that the
deflection of the string from the straight line vanishes at $\tau=0$.
The problem can be solved analytically but the result is given by
rather long expressions, so we do not reproduce it here.
Instead, we consider a limit when $X_0\to -\infty$.
To calculate the limit it is convenient to choose
$X_0=(\tilde{\tau}-\tau)\sinh\beta$ and, while keeping $\tilde{\tau}$
finite, to send $\tau\to \infty$. 
The physical interpretation of the new time-like parameter $\tilde{\tau}$
is quite simple---at $\tilde{\tau}=0$ the string crosses the $X=0$ 
plane\footnote{In the weak field approximation the motion of the string 
in the $X$-direction is not altered, i.e., $X=X_0+\tau\sinh\beta$.
Thus $\tau|_{X=0} = -X_0/\sinh\beta$.}.

 Omitting the terms vanishing
in this limit we obtain
\be
Z = \sigma + M\cosh\beta\ln\left(\frac{A_{-}B_{-}}
{A_{+}B_{+}}
 \right)\, ,
\ee
\[
Y = b - M\sinh\beta\left[\arctan\left(\frac{b^2+\tilde{\tau}(\tilde{\tau}+\sigma)\sinh^2\beta
}{b\sinh\beta\sqrt{{\tilde{\tau}}^2\sinh^2\beta+b^2+\sigma^2}}\right)+
\right.
\]
\vspace{0.3cm}
\[
+\arctan\left(\frac{b^2+\tilde{\tau}(\tilde{\tau}-\sigma)\sinh^2\beta}{b\sinh\beta\sqrt{
{\tilde{\tau}}^2\sinh^2\beta +b^2+\sigma^2}}\right)+
\]
\vspace{0.3cm}
\be
+\left.
\arctan\left(\frac{(\tilde{\tau}+\sigma)\sinh\beta}{b\cosh\beta}\right)
+\arctan\left(\frac{(\tilde{\tau}-\sigma)\sinh\beta}{b\cosh\beta} \right)
\right]\, ,
\ee
\be
X  = \tilde{\tau}\sinh\beta\, ,
\ee
\be
T = \tilde{\tau}\cosh\beta + M\ln\left(4\frac{B_{-}B_{+}}{A_{-}A_{+}}\right) 
+ 2M\ln(\tau\cosh^2\beta)\, ,
\ee
with
\be
A_\pm = b^2\cosh^2\beta+(\tilde{\tau}\pm\sigma)^2\sinh^2\beta \, , 
\ee
\be
B_\pm = \cosh\beta\sqrt{{\tilde{\tau}}^2\sinh^2\beta+b^2+\sigma^2} +
\tilde{\tau}\sinh^2\beta\pm \sigma \, .
\ee
The divergent term $2M\ln(\tau\cosh^2\beta)$ in the expression for 
$T$ arises because the
gravitational field at infinity falls down not rapidly enough.

It is easy to check that $Z$ is an antisymmetric function of $\sigma$, so
that for example $Z_{,\sigma\sigma}|_{\sigma=0}=0$.  To
obtain the condition of the coil formation one needs first to
calculate the derivatives $Z_{,\sigma}$ and $Z_{,\sigma\tilde{\tau}}$, and
then solve the equations
\be\n{C.1}
Z_{,\sigma}|_{\sigma=0}=0\, ,
\ee
\be\n{C.2}
Z_{,\sigma\tilde{\tau}}|_{\sigma=0}=0\, ,
\ee
in order to determine $\tau_*$ and $b$ as functions of rapidity
$\beta$. By calculating the derivatives we obtain
\be\n{C.3}
Z_{,\sigma}|_{\sigma=0}=
1-2M\cosh\beta\left(\frac{1}{\rho\cosh\beta + \tilde{\tau}\sinh^2\beta}
 +\frac{2\tilde{\tau}\sinh^2\beta}{{\tilde{\tau}}^2\sinh^2\beta+b^2\cosh^2\beta}   \right)
\, ,
\ee
\be\n{C.4}
Z_{,\sigma\tilde{\tau}}|_{\sigma=0}=
2M\cosh\beta \sinh^2 \beta \left(\frac{\tilde{\tau}\cosh\beta + \rho}
{(\rho\cosh\beta+\tilde{\tau}\sinh^2\beta)^2\rho} -
 \frac{2(b^2\cosh^2\beta-{\tilde{\tau}}^2\sinh^2\beta)}
{({\tilde{\tau}}^2\sinh^2\beta+b^2\cosh^2\beta)^2}
\right)
\, ,
\ee
where 
\be
\rho=\sqrt{{\tilde{\tau}}^2\sinh^2\beta+b^2}\, .
\ee
We use MAPLE to make all the above computations and to solve equations
(\ref{C.1})--(\ref{C.2}). The solution is
\be
b=2M\cosh\beta\, ,\hspace{1cm}
\tau_*=2M\cosh\beta\, .
\ee

The relation $b=2M\cosh\beta$ determines a coil formation curve. This
result  coincides with the one obtained earlier in  \cite{DVFr:99} as a
condition for coil formation for the ultrarelativistic
($\beta\to\infty$) motion of the string.

For comparison, we show the coil formation
curves for a non-rotating black hole in figure~\ref{coil}~(a) as 
calculated by using the weak field approximation.
The corresponding curve lies below the exact one.
One can conclude that the strong field effects help the coil
formation process.

\section{Discussions}
\label{discussion}
\setcounter{equation}0

In this paper we described the results of study of capture of a long
cosmic string by a rotating black hole and its critical
scattering. Since both problems are very non-linear, we use numerical
simulations. For capture and critical scattering of the string the
effects connected with the rotation of the black hole are very
profound. Partially it is connected with the fact that the dragging
into rotation effect increases the velocity of the central part of
the string  (as seen by an external observer) for prograde scattering
and decreases it for the retrograde scattering. Because it is easier to catch
a slower moving object the critical impact parameter
$b_c(v,a)$ is greater for retrograde motion than for the prograde one. We
calculated the critical impact parameter as a function of
velocity of the string and angular momentum of the black hole. These
plots have  interesting features for $v$  and $a/M$ close to 1.
In this regime the central part of the string can spend a considerable
 amount of time moving in the black hole vicinity before it gets captured or escapes.
 This makes such a regime of critical scattering
highly complicated. Since the system is non-linear, and there are two
qualitatively different final states (capture and scattering) one can
expect elements of chaotic behavior in this case. Such a chaotic
behavior occurs for example in the axisymmetric case when a circular
loop of a cosmic string moves in the Schwarzschild black hole metric
\cite{FrLa:99}.

One can also make the following observation. The internal geometry on
the world-sheet of the string is described by a time dependent
2-dimensional metric. When the string crosses the event horizon of
the bulk black hole, a region on the string surface is formed  from
where it is impossible to communicate with the parts of the string
which are outside the bulk black hole event horizon. In other words, a
two-dimensional black hole is created. The degrees of freedom living
on the string surface, e.g. transverse string perturbations,
propagate in this 2-dimensional spacetime with a 2-D black hole in
it \cite{FrLa:95,FrHeLa:96}. From this `2-D point of view' the scattering
of the string with the critical impact parameter is an event at the
threshold of the 2-D black hole formation. One can make a conjecture
that a  formation of a 2-D stringy hole obeys the scaling laws
similar to the universal scaling laws numerically discovered by
Choptuik \cite{Chop:93} in the general theory of relativity. A similar
effects for a world domain interacting with a black hole was discussed
in \cite{ChFrLa:98,FrLaCh:99}.

\vspace{12pt} 
{\bf Acknowledgments}:\ \ This work was partly supported by the Natural
Sciences and Engineering Research Council of Canada.  One of the
authors (V.F.) is grateful to the Killam Trust for its  financial
support. M.S. thanks  FS Chia PhD Scholarship. The authors are 
grateful to Don Page for various stimulating discussions.

\end{document}